%% file: TP2D.tex
\documentclass[pra,twocolumn,superscriptaddress]{revtex4-1}

\usepackage{graphicx}
\usepackage{color}
\usepackage[utf8]{inputenc}
\usepackage{amsmath}
\usepackage{verbatim}
\usepackage{amsbsy}
\newcommand{\fid}{\bar F}
\newcommand{\Tr}{\mathrm{Tr}}
\newcommand{\red}[1]{\textcolor{red}{#1}}

\newcommand{\E}{\mathcal{E}}
\newcommand{\rhoout}{\rho_{\text{out}}}
\newcommand{\rhoin}{\rho_{\text{in}}}
\newcommand{\Id}{\mathbb{1}}
\newcommand{\beq}{\begin{equation}}
\newcommand{\eeq}{\end{equation}} 
\newcommand{\bea}{\begin{eqnarray}}
\newcommand{\eea}{\end{eqnarray}}

\newcommand{\ketbra}[2]{|{#1}\rangle\langle{#2}|}
\newcommand{\ket}[1]{\left|{#1}\right\rangle}
\newcommand{\bra}[1]{\left\langle{#1}\right|}
\renewcommand\Re{\operatorname{Re}}
\renewcommand\Im{\operatorname{Im}}

\usepackage{qcircuit}

\input{macros.tex}

\begin{document}
\title{Selective and Efficient Quantum Process Tomography in Arbitrary Finite Dimension}

\author{Ignacio Perito}
\affiliation{Departamento de F\'isica, Facultad de Ciencias Exactas y Naturales, Universidad de 
Buenos Aires and IFIBA, CONICET, Ciudad Universitaria, 1428 Buenos Aires, Argentina.}
\author{Augusto Roncaglia}
\affiliation{Departamento de F\'isica, Facultad de Ciencias Exactas y Naturales, Universidad de 
Buenos Aires and IFIBA, CONICET, Ciudad Universitaria, 1428 Buenos Aires, Argentina.}
\author{Ariel Bendersky}
\affiliation{Departamento de 
Computación, Facultad de Ciencias Exactas y Naturales, Universidad de Buenos Aires and ICC, 
CONICET, Ciudad Universitaria, 1428 Buenos Aires, Argentina.}
\date{\today}
\begin{abstract}
 The characterization of quantum processes is a key tool in quantum information processing tasks 
for several reasons: on one hand, it allows to acknowledge errors in the implementations of 
quantum algorithms; on the other, it allows to charcaterize unknown processes occurring in Nature. 
In \cite{paper1conFer,paper2conFer} it was introduced a method to selectively and efficiently 
measure any given coefficient from the matrix description of a quantum channel. However, 
this method heavily relies on the construction of maximal sets of mutually unbiased 
bases (MUBs), which are  known to exist only when the dimension of the Hilbert space is the 
power of a prime number. In this article, we lift the requirement on the dimension by presenting 
two variations of the method that work on arbitrary finite dimensions: one uses tensor products of 
maximally sets of MUBs, and the other uses a dimensional cutoff of a higher prime power dimension.
\end{abstract}

\maketitle

\section{Introduction}

Being able to efficiently characterize the temporal evolution of a quantum system is one of the 
main tasks that is needed to accomplish in order to reliably operate  large-scale and general 
purpose quantum devices. In the last few years, there were proposed several methods for this issue, 
commonly known as quantum process tomography \cite{NielsenChuang, DCQD, MohsRezLid, paper1conFer, 
paper2conFer, paperConCeciLopez, prlChris, SCNQP}. In general, for an unknown evolution, full 
process tomography is 
inefficient since it requires resources that scale exponentially with the number of subsystems. 
However, it has been shown that is possible to efficiently obtain partial and relevant information 
about the process under consideration~\cite{paper1conFer, paper2conFer, paperConCeciLopez, prlChris, 
SCNQP}.

The temporal evolution of a quantum system can be described by a linear, completely positive, trace preserving map $\rhoout=\E\left(\rhoin\right)$. If the system under consideration has dimension $d$, one can choose a basis $\left\lbrace E_m, m=0...d^2-1\right\rbrace$ of operators and write the map as:
\begin{equation}
 \E\left(\rho\right)=\sum_{mn}\chi_{mn}E_m \rho E_n^\dagger
\label{eq:Mapa}
\end{equation}
where the trace preserving condition is given by $\sum_{mn}\chi_{mn}E_n^\dagger E_m=\Id$ and the matrix $\chi$ is Hermitian and positive, making the map completely positive.
In \cite{paper1conFer, paper2conFer, prlChris} the authors present a method that efficiently 
estimates  any coefficient from the $\chi$--matrix of a quantum channel. Their method relies on 
averaging the survival probability of a particular set of states, namely a state 2--design, when 
being acted upon by modified versions of the channel $\E$. 
The state 2--designs in this case are constructed using maximal sets of mutually unbiased bases 
(MUBs)~\cite{schwinger1960unitary}.
These bases are such that if a system is prepared in any state from one of them,
then a measurement with respect to other basis of this set  provides maximal uncertainty. The 
construction of state 2--designs imposes several limitations on the dimension $d$ for which this 
protocol can be implemented, since these maximal sets of bases are known to exist only for 
dimensions that are powers of prime 
numbers~\cite{ivanovic1981geometrical,wootters1989optimal,durt2010mutually}. For arbitrary 
dimensions only approximate constructions are known~\cite{Ambainis07}.

In this paper we present two variations of the above protocol that can be implemented in an 
arbitrary dimension. 
The first one is based on the construction of tensor products of  2--designs and measure the 
survival probability over modified channels. 
This idea not only allows to do selective and efficient  quantum process tomography in arbitrary systems, but also 
requires the preparation of product states in smaller dimensions.
%being less entangled than the states from a MUB set.  
The other scheme relies on the preparation of state 2--designs in a higher dimension and then 
project this design onto the desired dimension.

This paper is organized as follows. In Sec.~\ref{Sec:review}, we review some properties of 
state 2--designs, MUB sets and the results of Ref.~\cite{paper1conFer}, specifically, a protocol for 
selective and efficient quantum process tomography (SEQPT).  In Sec.~\ref{Sec:tensor}, we show that 
tensor products of 2--designs are approximate 2--designs up to local corrections, making them 
suitable for SEQPT; then we describe our first scheme. In Sec.~\ref{Sec:lowdim}, we introduce our 
second scheme, that performs SEQPT based on the construction of a state $2$--design on a higher 
dimension.

%xxxxxxxxxxxxxxxxxxxxxxxxxxxxxxxxxxxxxxxxxxxxxxxxxxxxxxxxxxxxxxxxxxxxxxxxxxxxxxxxxxxxxxxxxxxxxxxxxxxxxxxxxxxxxxxxxxxxxxxxxxxxxxxxxxxxxxxxxxxxxxxxxxxxxxxxxx
%xxxxxxxxxxxxxxxxxxxxxxxxxxxxxxxxxxxxxxxxxxxxxxxxxxxxxxxxxxxxxxxxxxxxxxxxxxxxxxxxxxxxxxxxxxxxxxxxxxxxxxxxxxxxxxxxxxxxxxxxxxxxxxxxxxxxxxxxxxxxxxxxxxxxxxxxxx

\section{SEQPT in prime power dimensions}
\label{Sec:review}
In this section we review the fundamental concepts for the protocol  developed in \cite{paper1conFer} that allows to perform selective and efficient quantum process tomography.
The method is based on the concept of state 2--designs, that allows to efficiently perform averages of quadratic functions over the Haar measure. 
Interestingly, the construction of 2--designs is intimately related to the existence of maximal sets mutually unbiased bases.

\subsection{2--designs and Mutually Unbiased Bases}

A finite set of states $\mathcal X=\left\lbrace\ket{\psi_m}, m=1,...,N\right\rbrace$ is a state 2--design if there is some fixed probability distribution $\{p_i\}_{i=1,\dots,N}$ such that
\begin{equation}
 \int_\mathcal H d\psi\,f\left(P_\psi\right)=\sum_{m=1}^N p_i f\left(P_{\psi_m}\right)
\end{equation}
for any $f$ that is quadratic in $P_\psi=\ketbra{\psi}{\psi}$, and the integral is made over the only normalized unitarily invariant measure on $\mathcal H$, namely, the Haar measure. 
That is, a state 2--design is a set of states such that the mean value of any quadratic function in 
$P_\psi$ over such a set gives the same mean value over the set of all possible states. When the probability distribution over the states in the set is uniform, we have $p_i=1/|\chi|=1/N$ for all $i=1,\dots,N$ and we refer to the set as a uniform 2--design. In the following, unless explictly noted, all 2--designs considered will be uniform. An interesting property that we will use extensively is that, for every pair of operators $A$ and $B$ in a Hilbert space of dimension $d$, it holds that
\begin{equation}\label{eq:traces}
 \left\langle A,B\right\rangle = \int_\mathcal H d\psi\, \Tr[P_\psi\,A\,P_{\psi}B]=\frac{\Tr A\; \Tr B+\Tr[AB]}{d(d+1)}.
\end{equation}

A simple way to find a state 2--design is to consider a set of mutually unbiased bases (MUB), which automatically form a state 2--design \cite{klappenecker2005mutually,dankert-2005}. 
Two orthonormal basis $\mathcal B_1$ and $\mathcal B_2$ of $\mathcal H$ are unbiased if 
every state of $\mathcal B_1$ has the same overlap with every state from $\mathcal B_2$ and viceversa. 
A set of orthonormal bases $\left\lbrace \mathcal B_1,...,\mathcal B_M\right\rbrace$ will be a mutually unbiased bases set if every pair of bases is unbiased. That is,
\begin{equation}
\left|\braket{\psi^J_l}{\psi^K_m}\right|^2=\delta_{JK}\delta_{lm}+\left(1-\delta_{JK}\right)\frac{1}{d},
\end{equation}
where $\ket{\psi^J_l}$ is the $l$--th state from basis $J$.
When the dimension $d$ is a power of a prime, the maximum number of mutually unbiased basis is $d+1$ \cite{ivanovic1981geometrical,wootters1989optimal}, 
and the $d(d+1)$ states form such set form a state 2--design \cite{klappenecker2005mutually}. 
For arbitrary dimension $d$ it is not known the maximum number of MUB.

\subsection{SEQPT}
\label{sec:seqpt}

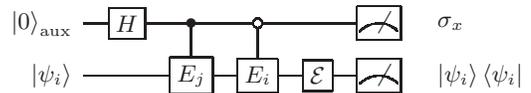
\begin{figure}
	\begin{equation*}
	\Qcircuit @C=1em @R=.7em {
		&\lstick{\ket{0}_{\text{aux}}}&\gate{H}&\ctrl{1}                          &\ctrlo{1}                    &\qw                          &\meter& \rstick{\sigma_x}\\
		&\lstick{\ket{\psi_i}}        & \qw    & \gate{E_j}       &\gate{E_i}   &\gate{\mathcal{E}}   &\meter&\rstick{\ket{\psi_i}\bra{\psi_i}}
	}
	\end{equation*}
	\caption{Circuit depicting the protocol for SEQPT. When averaging over all the states $\ket{\psi_i}$ from a state 2--design, the mean value of $\sigma_x$ ($\sigma_y$) conditioned to the survival of state $\ket{\psi_i}$ yields the real (imaginary) part of $\chi_{ij}$.}
	\label{seqpt}
\end{figure}

First, let us briefly describe the quantum algorithm for QPT introduced in Ref.~\cite{paper1conFer}. The method allows to efficiently obtain the elements of the 
$\chi$--matrix defined in Eq.~\eqref{eq:Mapa}, and can be easily understood if one considers an 
orthonormal operator basis $\{E_m\}$ satisfying $\Tr(E_mE_n^\dagger)=d\,\delta_{m,n}$ (orthonormality), 
 $E_mE_m^\dagger=\Id$ (unitarity), and  $E_0=\Id$. From  Eq.~\eqref{eq:traces} and the orthogonality of the operators, it is straightforward to verify that
 the mean fidelity or survival probability is given by:
\beq
\bar F(\mathcal E)=\int_\mathcal H d\psi \; \Tr[P_\psi \, \E \left(P_{\psi}\right)],
\eeq 
and provides information about the diagonal contribution of $E_0$. In general, each
element $\chi_{ij}$ can be obtained by measuring the following averaged survival probability:
%$$
% \int d\psi \bra{\psi}\mathcal E \left( \ket{\psi}\bra{\psi} \right)\ket{\psi} = \frac{D\chi_{00}+1}{D+1}.
%$$
\beq
\label{eq:meanfid}
\bar F(\E_{ij})= \frac{d\,\chi_{ij}+\delta_{ij}}{d+1}.
\eeq
where  $\E_{ij}(\rho)\equiv\E(E^\dagger_i \rho E_j)$.
Thus, using a slightly modified channel, it would be possible 
to measure the elements of the $\chi$--matrix provided
one is able to perform the average over the Haar measure.
%Therefore, one can perform diagonal tomography by measuring the average survival property of every state of the Hilbert space over the channel (or the modified channel). 
The key point of the method is that, since both expressions are quadratic in $P_\psi$, this average can be evaluated by considering just a finite set of states from a $2$--design.
Notice that the modified channel $\E_{ij}$ is not physical, however, it can be implemented via the 
tomographic protocol depicted in Fig. \ref{seqpt}. 
Then, it is straightforward to show that the real (imaginary) part of $\chi_{ij}$ is obtained by measuring the mean value of $\sigma_x$ ($\sigma_y$) conditioned on the survival 
of the state $\ket{\psi_i}$, where $\ket{\psi_i}$ is randomly chosen from a state 2--design.

%xxxxxxxxxxxxxxxxxxxxxxxxxxxxxxxxxxxxxxxxxxxxxxxxxxxxxxxxxxxxxxxxxxxxxxxxxxxxxxxxxxxxxxxxxxxxxxxxxxxxxxxxxxxxxxxxxxxxxxxxxxxxxxxxxxxxxxxxxxxxxxxxxxxxxxxxxx
%xxxxxxxxxxxxxxxxxxxxxxxxxxxxxxxxxxxxxxxxxxxxxxxxxxxxxxxxxxxxxxxxxxxxxxxxxxxxxxxxxxxxxxxxxxxxxxxxxxxxxxxxxxxxxxxxxxxxxxxxxxxxxxxxxxxxxxxxxxxxxxxxxxxxxxxxxx

\section{General SEQTP with tensor products of 2--designs}
\label{Sec:tensor}
\subsection{Tensor products of 2--designs}

In this Section we will first describe how tensor products of 
2--designs can be used to approximate 2--designs,
and then we show that tensor products of maximally MUB sets provide a good approximation for 
integration purposes.

Let us start with the bipartite case. Consider a Hilbert space $\mathcal{H}=\mathcal{H}_1\otimes \mathcal{H}_2$ of dimension $d=D_1D_2=p_1^{n_1}p_2^{n_2}$, with $p_1$ and $p_2$ different prime numbers.
This ensures that a maximally set of MUBs exists for both dimensions $D_1$  and $D_2$, and that those MUB sets are $2$--designs.
Let $A$ and $B$ be operators in dimension $d$, and consider the following integral
\begin{equation}
\label{eq:YAB}
\left \langle A,B \right \rangle_\otimes=\int_{\mathcal{H}_1}\int_{\mathcal{H}_2} d\psi_1\, d\psi_2 \; \Tr[P_{\psi_1 \psi_2}\, A\,P_{\psi_1 \psi_2} \,B],
\end{equation}
where $P_{\psi_1 \psi_2}=\ketbra{\psi_1}{\psi_1}\otimes \ketbra{\psi_2}{\psi_2}$ and the integrations are performed over the Haar measure for $\mathcal{H}_1$ and $\mathcal{H}_2$ respectively. 
It is easy to see that the above expression can be evaluated using 2--designs in $\mathcal{H}_1$ and $\mathcal{H}_2$:
$\left \langle A,B \right \rangle_\otimes=\frac{1}{|X_\otimes|}\sum_{\psi\in X_\otimes}\Tr[P_{\psi}\, A\,P_{\psi} \,B]$,
where the sum is taken over the product states $\ket\psi \equiv \ket{\psi_1}\ket{\psi_2}\in X_\otimes \equiv X_1 \otimes X_2$ with $\ket{\psi_1}\in X_1$ 
and $\ket{\psi_2}\in X_2$,  $X_1$ and $X_2$ are $2$--designs in $\mathcal{H}_1$ and $\mathcal{H}_2$ respectively.  
Now, expanding the operators $A$ and $B$ in a product basis 
and using the identity of Eq.~\eqref{eq:traces} for each dimension $D_i$  we arrive at:
\bea\label{dosDisenhoDosPartes}
&& \left \langle A,B \right \rangle_\otimes=\frac{1}{d(D_1+1)(D_2+1)}\big(\Tr A\, \Tr B+ \Tr (AB)+\nonumber \\
&&\quad \quad + \Tr\left[ A \,(\mathbb{I}_1\otimes \Tr_1 (B))  \right]+\Tr\left[ A\, ( \Tr_2 (B)\otimes \mathbb{I}_2)  \right]\big)
\eea
where $\mathbb{I}_i$ is the identity over system $i$ and $ \Tr_i$ is the partial trace over system $i$.
Notice that the first two terms in Eq.~\eqref{dosDisenhoDosPartes} are proportional to the ones that would appear in averages over a $2$--design (see Eq.~\eqref{eq:traces}).
In fact, for an average like $\left \langle A,B \right \rangle_\otimes$, one can use two 
$2$--designs and the resulting 
expression would be similar to Eq.~\eqref{eq:traces}, with the addition of two new terms. Thus,
\bea \label{eq:HaarDosPartes}
&& \left \langle A,B \right \rangle= \frac{\left(D_1+1\right)\left(D_2+1\right)}{(d+1)}\left \langle A,B \right \rangle_\otimes + \nonumber \\
&& \quad \quad-\frac{\Tr\left[ A\, (\mathbb{I}_1\otimes \Tr_1 (B))  \right] + \Tr\left[ A\, ( \Tr_2 (B)\otimes \mathbb{I}_2)  \right]}{d\,(d+1)}.
\eea
We will see later, in the context of QPT, that the last two terms can either be neglected (depending on the values of $D_i$) or measured.
Thus, Eq.~\eqref{eq:HaarDosPartes} establishes  the connection between averages over 2-designs $\left \langle A,B \right \rangle$ and averages over tensor 
product of 2-designs  $\left \langle A,B \right \rangle_\otimes$ in the bipartite case.

In general, when the dimension $d$ is the product of more than two prime power factors  
$d=p_1^{n_1}...p_N^{n_N}$, 
we will consider tensor products of $N$ different 2--designs. In this case, one can show that the following identity holds:
\begin{equation}
\left \langle A,B \right \rangle_\otimes=\sum_{X\in\left\lbrace 0,1\right\rbrace^N} \frac{\Tr \left[ A \left( \Id^{(X)}\otimes \Tr_X B \right) \right]}{d\prod_{k=1}^N (D_k+1)},
\end{equation}
where $\mathbb{I}^{(X)}$ is the identity operator on the subsystems for which the binary $N$--uple $X$ has a $1$, and $\Tr_X$ is the partial trace over those subsystems.
Hence, this identity is the extension of Eq.~\eqref{dosDisenhoDosPartes} when the average is taken 
over general tensor products of 2--designs.

%xxxxxxxxxxxxxxxxxxxxxxxxxxxxxxxxxxxxxxxxxxxxxxxxxxxxxxxxxxxxxxxx

\subsection{SEQPT with tensor products of $2$-designs: the bipartite case}
\label{Sec:seqptbip}
\subsubsection{Mean survival probabilities}

%Let us start with the simplest case where the dimension $d$ is such 
Consider that $d=D_1 D_2$ with $D_1=p_1^{n_1}$ and $D_2=p_2^{n_2}$.
%where $p_i$ are prime numbers.
 As it was shown in Sec.~\ref{sec:seqpt}, the original method relies on the fact that one can efficiently perform averages,
 like the one in Eq.~\eqref{eq:meanfid},  when the dimension of the system is a prime power. Here we will show that
this could also be done by taking averages over tensor product of 2-designs.
First, let us expand the channel of Eq.~\eqref{eq:Mapa} in a product basis of operators acting on $\mathcal{H} = \mathcal{H}_1 \otimes \mathcal{H}_2$.
This basis can be written in terms of two orthogonal basis of each system: $\{ E_{j_1j_2} \equiv E_{j_1} \otimes E_{j_2} \}_{j_1=0,\dots,D_1^2-1}^{j_2=0,\dots,D_2^2-1}$,
where each element $E_{j_i}$ is a unitary matrix.
Thus, the channel in this basis is defined as:
\begin{equation}
  \E(\rho) =  
\sum_{\mu_1\mu_2\nu_1\nu_2} \chi_{\nu_1\nu_2}^{\mu_1\mu_2} E_{\mu_1\mu_2} \rho   E^{\nu_1\nu_2}   
\, ,
\label{canal2primos}
\end{equation}
for some coefficients $\chi_{\nu_1\nu_2}^{\mu_1\mu_2}$. We have adopted the convention $E^{j_1j_2} \equiv E_{j_1j_2}^\dagger$ and $E^{j_i}\equiv E_{j_i}^\dagger$,
and from now on we will also consider $\delta_i^j\equiv\delta_{ij}$.

In order to obtain an arbitrary coefficient $\chi_{i_1i_2}^{j_1j_2}$,
we can consider the modified channel $\E_{j_1j_2}^{i_1i_2} (\rho) \equiv \E \left( E^{i_1i_2} \rho E_{j_1j_2} \right)$.
Using this modified channel, the averaged  survival probability is equal to:
\beq
\fid(\E_{j_1j_2}^{i_1i_2})= \frac{ d\,\chi^{i_1i_2}_{j_1j_2} + \delta^{i_1}_{j_1}\delta^{i_2}_{j_2} }{d+1}.
\label{eq:fid}
\eeq
Which is useful provided one can implement the average using state 2-designs but, as we 
mentioned before, this is not the case for an arbitrary dimension $d$.
So let us consider the following mean survival probability:
\beq 
\bar F_\otimes(\E_{j_1j_2}^{i_1i_2})=\int_{\mathcal H_1} \int_{\mathcal H_2} d\psi_1 d\psi_2\,\Tr[P_{\psi_1\psi_2} \,\E_{j_1j_2}^{i_1i_2} \left(P_{\psi_1\psi_2}\right)].
\nonumber
\eeq 
Notice that this expression can be evaluated by averaging over a finite set of states belonging
to $X_\otimes$. Additionally, one can show that the elements of the $\chi$-matrix 
obey  the following identity (see Appendix \ref{app:tensor}):
\bea
\label{eq:chifid}
&&\chi_{j_1j_2}^{i_1i_2}=\fid_\otimes(\E_{j_1j_2}^{i_1i_2})\frac{(1+D_1)(1+D_2)}{d}+\frac{\delta^{
i_1}_{j_1}\delta^{i_2}_{j_2} }{d}\nonumber \\
&&\quad \quad 
-\fid_1(\E_{j_1j_2}^{i_1i_2})\frac{(1+D_1)}{d}-\fid_2(\E_{j_1j_2}^{i_1i_2})\frac{(1+D_2)}{d}.
\eea
The last terms in the expression have a clear interpretation: $\fid_1$ can be thought as a reduced mean survival fidelity over system 1:
$\fid_1(\E_{j_1j_2}^{i_1i_2})=\frac{1}{d_2|X_1|}\sum_{\psi_1\in X_1}\Tr\left[(P_{\psi_1}\otimes\Id_2) \,\E_{j_1j_2}^{i_1i_2}(P_{\psi_1}\otimes\Id_2)]\right]$.~In terms of integrals over the Haar measure it can  be written as: 
${\fid_1(\E_{j_1j_2}^{i_1i_2})=\int_{\mathcal{H}_1} d\psi_1 \bra{\psi_1}\Tr_2(\E_{j_1j_2}^{i_1i_2}(P_{\psi_1}\otimes\Id_2/d_2))\ket{\psi_1}}$,
and similarly for $\fid_2(\E_{j_1j_2}^{i_1i_2})$.
Thus, the above identity tells us that we do not need to compute Haar integrals in order to evaluate the elements  $\chi_{j_1j_2}^{i_1i_2}$,
it is enough to perform a sampling over a finite set of states from $X_\otimes$. 
Notice also that the first term in the right hand side of the last equation is the mean survival probability of the elements of our 
approximate $2$--design over a channel that is non-physical.  This feature already appears in the 
prime power dimension case, and how to circumvent this issue was described in Ref.~\cite{paper1conFer}. 
Bellow, we describe how to determine each of the terms with a single quantum circuit.

\subsubsection{Estimation of the elements of the $\chi$-matrix}

\begin{figure}
	\begin{equation*}
	  \Qcircuit @C=1em @R=.7em {
	  	&\lstick{\ket{0}_{\text{aux}}}&\gate{H}&\ctrl{1}                          &\ctrlo{1}                    &\qw                          &\meter& \rstick{\sigma_x}\\
	  	&\lstick{\ket{\psi_1}}        & \qw    & \multigate{1}{E^{i_1 i_2}}       &\multigate{1}{E^{j_1 j_2}}   &\multigate{1}{\mathcal{E}}   &\meter&\rstick{\ket{\psi_1}\bra{\psi_1}}\\
	  	&\lstick{\ket{\psi_2}}        & \qw    &\ghost{E^{i_1 i_2}}       &\ghost{E^{j_1 j_2}}   &\ghost{\mathcal{E}}   &\meter&\rstick{\ket{\psi_1}\bra{\psi_1}}
	  }
	\end{equation*}
	\caption{Circuit that estimates the real part of $\chi^{i_1i_2}_{j_1j_2}$. 
		The initial states $\ket {\psi_1}$ and $\ket {\psi_2}$ are uniformly sampled from the set of states 
		that belongs to the 2-designs $X_1$ and $X_2$ respectively.
		A measurement of the expectation value of $\sigma_x$ ($\sigma_y$) conditioned to the results of the measurements at the other two systems allows  
		to estimate the $\Re(\chi^{i_1i_2}_{j_1j_2})$ ($\Im(\chi^{i_1i_2}_{j_1j_2})$). }
	\label{seqpt2primos}
\end{figure}
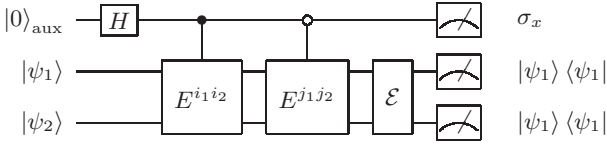

In order to estimate the coefficients $\chi^{i_1i_2}_{j_1j_2}$ we have to determine three different complex terms, as it is shown in Eq.~\eqref{eq:chifid}.
These terms can be written as mean survival probabilities over non-physical quantum channels. The real part of these terms can be 
estimated through the circuit represented in Fig.~\ref{seqpt2primos}. This is a variation of the strategy used in Ref. \cite{paper1conFer}.
The method utilizes one auxiliary qubit whose polarization is measured. 
The inputs of the circuit are the clean qubit and states randomly chosen from each 2-design. It is 
easy to show that by measuring the expectation value of $\sigma_x\otimes P_{\psi_1} 
\otimes P_{\psi_2}$
one estimates $\Re(\fid_\otimes(\E_{j_1j_2}^{i_1i_2}))$. Analogously, with the expectation value of $\sigma_y\otimes P_{\psi_1} \otimes P_{\psi_2}$
one estimates  $\Im(\fid_\otimes(\E_{j_1j_2}^{i_1i_2}))$. 

The other terms, $\fid_1(\E_{j_1j_2}^{i_1i_2})$ and $\fid_2(\E_{j_1j_2}^{i_1i_2})$, can also be estimated
with the same circuit.  Thus, for instance, in order to estimate
$\Re(\fid_1(\E_{j_1j_2}^{i_1i_2}))$ one has to measure the expectation of $\sigma_x\otimes 
P_{\psi_1} \otimes \Id_2$ given  the initial state 
$\ketbra{0}{0}_{\rm aux}\otimes P_{\psi_1} \otimes \frac{\Id_2}{D_2}$. This is achieved by looking at 
the statistics of the measurements of the auxiliary qubit and system 1
independently from the results of the measurements at system 2.
This is so because the initial state of system 2 is a random state from $(1+D_2)$ orthogonal bases, which is a possible implementation of $\Id_2$ provided
the result of the measurement of system 2 is not taken into account.
Thus, each run of the circuit in Fig.~\ref{seqpt2primos} yields one of eight possible outputs for each polarization measurement. The number of outputs is the result of the product of
2 values for the polarization, 2 values for $P_{\psi_1}$ and 2 values for $P_{\psi_1}$. In summary, detection of $P_{\psi_1}$ along with $P_{\psi_2}$ contributes 
to the estimation of $\fid_\otimes$, detection of just $P_{\psi_1}$ to the estimation of $\fid_1$, and detection of just $P_{\psi_2}$ to the estimation of $\fid_2$. 

%\begin{figure}
% \includegraphics[width=0.48\textwidth]{seqpt2primos.png}
% \caption{Circuit that estimates the real part of $\chi^{i_1i_2}_{j_1j_2}$. 
% The initial states $\ket {\psi_1}$ and $\ket {\psi_2}$ are uniformly sampled from the set of states 
%that belongs to the 2-designs $X_1$ and $X_2$ respectively.
%A measurement of the expectation value of $\sigma_x$ ($\sigma_y$) conditioned to the results of the measurements at the other two systems allows  
%to estimate the $\Re(\chi^{i_1i_2}_{j_1j_2})$ ($\Im(\chi^{i_1i_2}_{j_1j_2})$). }
% %\label{seqpt2primos}
%\end{figure}

So far this protocol seems to be inefficient: given that 
$|X_\otimes|$ grows exponentially with $n_1$ and $n_2$, one would expect that also the number of experiments required to estimate 
$\Re(\fid_\otimes(\E_{j_1j_2}^{i_1i_2}))$  grows exponentially with $n_1$ and $n_2$. Fortunately, will show that, altough the number of experiments
  required to estimate an element of the $\chi$-matrix depends upon the 
desired precision, it is independent of $n_1$ and $n_2$. This can easily be understood by looking at Eq.~(\ref{eq:chifid}) and 
realising that $\Re \left( \chi_{j_1j_2}^{i_1i_2} \right)$ can be thought of as an average of a random variable 
which can take eight values. Then, one can relate the number of trials needed to estimate 
this average, $M$,  with the desired error $\epsilon$ and failure probability $p$ as (see Appendix \ref{app:efficiency}):
\begin{equation}
\label{efficiency2}
p \leq 2 \exp \left( \frac{-M\epsilon^2}{2(1-\frac{1}{d})^2} \right).
\end{equation}
This implies the estimation of an element of the $\chi$-matrix requires a number of experiments of 
the order $M\geq{2 \ln (2/p)}/{\epsilon^2}$. 
%Therefore, this number of experiments is enough to estimate $\Re{\left(\chi^{i_1i_2}_{j_1j_2}\right)}$ with error $\epsilon$ and probability of failure lower than $p$.}

%xxxxxxxxxxxxxxxxxxxxxxxxxxxxxxxxxxxxxxxxxxxxxxxxxxxxxxxxxxxxxxxx
\subsection{SEQPT with tensor products of $2$-designs: the general case}

Now we consider the tensor product approach for a system of arbitrary dimension $d$. In this case,
we can always factorize it as $d=\prod_{a=1}^N p_a^{n_a}$ where 
$p_a$ are prime numbers and $n_a \geq 1$ their respective multiplicity in the prime 
number factorization of $d$.  Thus, the Hilbert space of the system is $\mathcal{H} = \bigotimes_{a=1}^N \mathcal{H}_a$ 
and each subspace has dimension $D_a = p_a^{n_a}$. %The analysis presented in the previous subsection
%corresponds to $N=2$. 
As it was done before, we will consider unitary orthogonal operator bases for each subsystem
$\left\{ E_{i_a}^a \right\}_{i_a = 1,\dots,D_a^2}$, and 
the basis of operators for the composite system as $\left\{ \bigotimes _{a=1}^N E_{i_a}^a  \equiv E_{i}\right\}$, 
where from now on we will assume that $i\equiv i_1...i_N$. % \equiv E_{i_1 \dots i_N}\right\}$. 
%In the following, we will adopt the convention of summing over repeated indices. 
Thus, an arbitrary channel $\E$ can be expanded as:   
\beq
% \E(\rho) = \chi_{\nu_1\dots \nu_N}^{\mu_1\dots \mu_N} E_{\mu_1\dots \mu_N } \rho 
%E^{\nu_1\dots \nu_N } \, ,
\E(\rho) = \sum_{ \nu, \mu}\, \chi_{  \nu}^{ \mu}\; E_{ \mu}\, \rho\, E^{ \nu} \,.
\eeq

\begin{figure}
	\begin{equation*}
	\Qcircuit @C=1em @R=.7em {
		&\lstick{\ket{0}_{\text{aux}}}&\gate{H}&\ctrl{1}                          &\ctrlo{1}                    &\qw                          &\meter& \rstick{\sigma_x}\\
		&\lstick{\ket{\psi_1}}        & \qw    & \multigate{3}{E^{i}}       &\multigate{3}{E^{j}}   &\multigate{3}{\mathcal{E}}   &\meter&\rstick{\ket{\psi_1}\bra{\psi_1}}\\
		&\lstick{\ket{\psi_2}}        & \qw    &\ghost{E^{i}}       &\ghost{E^{j}}   &\ghost{\mathcal{E}}   &\meter&\rstick{\ket{\psi_2}\bra{\psi_2}}\\
		&\lstick{     \vdots       }  &        &\ghost{E^{i}}       &\ghost{E^{j}}   &\ghost{\mathcal{E}}   & \vdots     &\\
		&\lstick{\ket{\psi_N}}        & \qw    &\ghost{E^{i}}       &\ghost{E^{j}}   &\ghost{\mathcal{E}}   &\meter&\rstick{\ket{\psi_N}\bra{\psi_N}}
	}
	\end{equation*}
	\caption{Circuit that estimates the real part of $\chi^{i}_j$ for a process that acts on a system of arbitrary dimension $d=\prod_{a=1}^N p_a^{n_a}$. 
		The initial states $\ket {\psi_i}$ are uniformly sampled from the set of states that belongs to the 
		2-designs $X_1,\dots ,X_N$.
		A measurement of the expectation value of $\sigma_x$ ($\sigma_y$) conditioned to the results of the measurements at the other systems allows  
		to estimate the $\Re(\chi^{i}_j)$ ($\Im(\chi^{i}_j)$). }
	\label{fig:gencirc}
\end{figure}
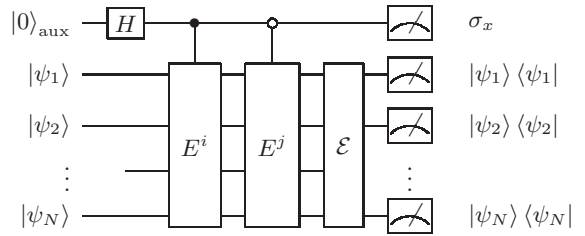

Our objective is to estimate effciently an arbitrary coefficient $\chi_{j}^{i}$.  %$\chi_{j_1\dots j_N}^{i_1\dots i_N}$.  
The crucial objects we need to compute are integrals of the form $\left \langle M,Q \right \rangle=\int_\mathcal{H} \, d\psi \,\Tr [P_\psi M P_\psi Q]$.
 Let us consider the $2$-designs $X_a \in \mathcal{H}_a$ for $a=1,\dots,N$, which we know how to construct because each 
$\mathcal{H}_a$ has prime power dimension. Now we can define our approximate $2$-design $X_\otimes \in 
\mathcal{H}$ as  $X_\otimes = \bigotimes_{a=1}^N  X_a$. It is easy to check that, 
if $M = \bigotimes_{a=1}^N M_a$ and $Q = \bigotimes_{a=1}^N Q_a$:
\begin{equation}
\left \langle M,Q \right \rangle_\otimes= \prod_{a=1}^N 
\left( \Tr M_a \, \Tr Q_a + \Tr[M_aQ_a] \right).
\label{sumaNfactores}
\end{equation}
If we expand  the right hand side of the last equation, we obtain $2^N$ terms. A 
convenient way to refer to each of those terms is the following:
\begin{equation}
 f_{MQ} (m_1,\dots,m_N) \equiv \prod_{a=1}^N \Tr[M_a Q_a]^{m_a \oplus 1} (\Tr M_a\,\Tr Q_a)^{m_a}, \nonumber
\end{equation}
where $m_1 \dots m_N$ take values in $\{0,1\}$ and $\oplus$ is the addition modulo-$2$. Note that 
the sum of two of these terms is given by:
\beq
f_{MQ} (0,\dots,0) +  f_{MQ} (1,\dots,1)  =  d(d+1)  \langle M,Q \rangle.
\eeq
Therefore, the integral we are interested in appears in equation 
Eq.~\eqref{sumaNfactores}, but we have to deal with some extra terms in order perform the process tomography. 
We will show that the circuit for the estimation of non-diagonal coefficients is analogous to the one presented for $N=2$ and it is shown in Fig.~\ref{fig:gencirc}.

The procedure we can use to obtain the elements of the $\chi$-matrix in terms of mean survival probabilities is analogous as the 
one we follow in the bipartite case (see Appendix \ref{app:tensor}). Thus, one can show that the elements of the $\chi$-matrix obey:
\bea
\label{eq:chifidN}
%&& \chi_{j_1..j_N}^{i_1...i_N}=\fid_\otimes(\E_{j_1..j_N}^{i_1...i_N})\,\frac{\prod_{a=1}^{N}(D_a+1)}{d}+\frac{\prod_{a=1}^N \delta_{j_a}^{i_a}}{d} \nonumber \\
%&&\quad \quad-\sum_{\vec m\neq\{\vec 0, \vec 1\}}  \fid_{\vec m}(\E_{j_1...j_N}^{i_1...i_N}) \; \frac{\prod_{a/m_a=1} (D_a+1)}{d}
&& 
\chi_{j}^{i}=\fid_\otimes(\E_{j}^{i})\,\frac{\prod_{a=1}^{N}(D_a+1)}{d}+\frac{(2^N 
-3)\prod_{a=1} ^N \delta_{j_a}^{i_a}}{d} \nonumber \\
&&\quad \quad-\sum_{\vec m\neq\{\vec 0, \vec 1\}}  \fid_{\vec m}(\E_{j}^{i}) \; \frac{\prod_{a/m_a=1} (D_a+1)}{d},
\eea
where $\vec{m}$ is a vector with $N$ components each of which take values in $\{0,1\}$.  
In this case, the reduced mean survival probabilities are defined as:
\beq
%\fid_{\vec m}(\E_{j_1...j_N}^{i_1...i_N})=\frac{\prod_{a/m_a=0} (D_a+1)}{|X_\otimes|} \sum_{\psi \in X_{\vec m}}
%\Tr [ \rho_{\psi}^{\vec{m}} \,\E_{j_1...j_N}^{i_1...i_N}( \rho_{\psi}^{\vec{m}}) ] \nonumber
\fid_{\vec m}(\E_{j}^{i})=\frac{\prod_{a/m_a=0} (D_a+1)}{|X_\otimes|} \sum_{\psi \in X_{\vec m}}
\Tr [ \rho_{\psi}^{\vec{m}} \,\E_{j}^{i}( \rho_{\psi}^{\vec{m}}) ], \nonumber
\eeq
where the averages are taken over the states in ${X_{\vec{m}} \equiv \bigotimes_{a / \vec{m}_a=1} 
X_a}$, and states defined as:
\begin{equation}
 \rho_{\psi}^{\vec{m}} = \left( \bigotimes_{a / m_a=0}  \frac{\Id^{(a)}}{D_a} \right) 
\otimes 
P_\psi \,\, \text{with} \,\, \psi \in  X_{\vec m}.
\end{equation}
These are completely mixed states in the subsystems labelled by $m_a=0$, and elements of a 
$2$-design  for 
the rest.

The circuit that estimates the elements of the $\chi$-matrix is shown in Fig.~\ref{fig:gencirc}. There, as in the bipartite case, 
the real and imaginary part of the fidelities are estimated by measuring mean values associated to $\sigma_x$ and $\sigma_y$,
respectively. Furthermore, all the terms in Eq.~\eqref{eq:chifidN} can be evaluated by considering 
many runs of the same circuit, where the initial states 
$\ket{ \psi_a}$ are sampled from the state 2-design in $\mathcal H_a$. In this case, the circuit has $2^{N+1}$ outputs: 2 for the auxiliary qubit, and 
2 for each of the $N$ subsystems (the projective measurement gives binary results).  Thus, one can discriminate between these results to evaluate all these fidelities.
The efficiency of this protocol is 
discussed in Appendix \ref{app:efficiency}.

%\begin{figure}
% \includegraphics[width=0.48\textwidth]{GenCirc.JPG}
% \caption{Circuit that estimates the real part of $\chi^{i}_j$ for a process that acts on a system of arbitrary dimension $d=\prod_{a=1}^N p_a^{n_a}$. 
% The initial states $\ket {\psi_i}$ are uniformly sampled from the set of states that belongs to the 
%2-designs $X_1,\dots ,X_N$.
%A measurement of the expectation value of $\sigma_x$ ($\sigma_y$) conditioned to the results of the measurements at the other systems allows  
%to estimate the $\Re(\chi^{i}_j)$ ($\Im(\chi^{i}_j)$). }
% %\label{fig:gencirc}
%\end{figure}

%xxxxxxxxxxxxxxxxxxxxxxxxxxxxxxxxxxxxxxxxxxxxxxxxxxxxxxxxxxxxxxxxxxxxxxxxxxxxxxxxxxxxxxxxxxxxxxxxxxxxxxxxxxxxxxxxxxxxxxxxxxxxxxxxxxxxxxxxxxxxxxxxxxxxxxxxxx
%xxxxxxxxxxxxxxxxxxxxxxxxxxxxxxxxxxxxxxxxxxxxxxxxxxxxxxxxxxxxxxxxxxxxxxxxxxxxxxxxxxxxxxxxxxxxxxxxxxxxxxxxxxxxxxxxxxxxxxxxxxxxxxxxxxxxxxxxxxxxxxxxxxxxxxxxxx

\section{General SEQPT using low dimensional projection}
\label{Sec:lowdim}

In this section we show a different strategy that allows to  construct a non-uniform
$2$--design in any dimension $d$ different from a prime power. 
The  method is based on taking a 2--design in a dimension grater than $d$, and then 
project this set over dimension $d$.

Let us consider $D$ an integer such that ${D>d}$ and ${D=p^n}$ for some prime number $p$  and natural number $n$. 
$D$ could be, for instance, the smallest dimension grater than $d$ for which we know how to 
generate $2$--designs.
We assume that in dimension $D$ a maximal MUB set is made up of the bases $\mathcal B_0, ..., \mathcal B_D$.
The basis $\mathcal B_0=\left\lbrace \ket{\psi^0_0},...,\ket{\psi^0_{D-1}}\right\rbrace$ can be arbitrary chosen, and the sates from the other bases of the MUB set can be written as
\beq
 \ket{\psi^M_k} = \sum_{j=0}^{D-1} \left. \frac{e^{i\alpha^M_{jk}}}{\sqrt{D}} \ket{\psi^0_j} \right., 
\eeq
where $\ket{\psi^M_j}$ denotes the $j$--th state from basis $\mathcal B_M$ \cite{wootters1989optimal}. 
The set of states $X_D= \left\lbrace \ket{\psi^M_j}\right\rbrace^{M=0,...,D}_{j=0,...,D-1}$ form a state $2$--design in dimension $D$ \cite{klappenecker2005mutually}.

The key point  of the method is, as before, the efficient estimation of Haar integrals of quadratic 
forms.
Given two arbitrary operators $A$ and $B$ in $\mathcal{H}_d$, and their expansion in an orthonormal basis $\{ \ket{j} 
\}_{j=0,\dots,d-1}$:
\begin{equation}
 A = \sum_{j,k=0}^{d-1}a_{jk}\ket{j}\bra{k} \,\, , \,\, 
 B = \sum_{j,k=0}^{d-1}b_{jk}\ket{j}\bra{k} \, ,  
\end{equation}
we can define two operators that can be considered as their extensions to $\mathcal{H}_D$:
\begin{equation}
 \tilde{A} \equiv \sum_{j,k=0}^{d-1}a_{jk}\ket{\psi_j^0}\bra{\psi_k^0} \,\, , \,\, 
 \tilde{B} \equiv \sum_{j,k=0}^{d-1}b_{jk}\ket{\psi_j^0}\bra{\psi_k^0} \, .  
\end{equation}
Then, as $X_D$ is a $2$-design in $\mathcal{H}_D$, we have that:
\beq
 \frac{1}{|X_D|} \sum_{j,M} \Tr[\tilde{A}P_{\psi_j^M}] \, \Tr[\tilde{B}P_{\psi_j^M}] \, = \frac{\Tr\tilde{A}\, \Tr\tilde{B}+\Tr[\tilde{A}\tilde{B}]}{D(D+1)} 
\eeq
with $P_{\psi_j^M}=\ketbra{\psi_j^M}{\psi_j^M}$. 
We can define the projector ${\Pi^D_d=\sum_{j=0}^{d-1} \ket{\psi^0_j}\bra{\psi^0_j}}$ and the following set 
 of unnormalized vectors $\left\{\ket{\tilde{\phi}_j^M} \equiv \Pi_d^D \ket{\psi_j^M}\right\}^{M=0,...,D}_{j=0,...,D-1}$.
Since both $\tilde{A}$ and $\tilde{B}$ act on the subspace spanned by the first $d$ 
elements of $\mathcal B_0$, they satisfy $\tilde{A} = \Pi_d^D \tilde{A}\, \Pi_d^D$ and 
$\tilde{B} = \Pi_d^D \tilde{B} \,\Pi_d^D$, then:
\bea
\Tr[\tilde{A}\, P_{\psi_j^M}] = \bra{\tilde{\phi}_j^M} \tilde{A} \ket{\tilde{\phi}_j^M}\nonumber \\
\Tr[\tilde{B}\, P_{\psi_j^M}] = \bra{\tilde{\phi}_j^M} \tilde{B} \ket{\tilde{\phi}_j^M}.
\eea

Now, let us consider the set  $X_d ^D=\left\{ \ket{\phi^M_j}\right\} \in \mathcal{H}_d$ defined
by the following states:
\begin{equation}
 \ket{{\phi}_j^M} = \begin{cases}
                           \ket{j} \,\, & \text{if}  \,\, M=0 \,\, \text{and} \,\,0\leq  j\leq 
d-1\\
%                           0  \,\, & \text{if}  \,\, M=0 \,\, \text{and} \,\, j\geq  d\\

 \sum_{k=0}^{d-1} \frac{e^{i\alpha_{kj}^M}}{\sqrt{d}} \ket{k}  \,\, & \text{if}  \,\, 1\leq M \leq (D+1) \, \\
& \text{and} \,\,0\leq j \leq D-1\, .
 
                           \end{cases} \nonumber
\end{equation}
This set is composed by $|X_d ^D|=d+D^2$ states, and it is easy to check that:
\begin{equation}
\Tr[\tilde{A}\, P_{\psi_j^M}]= \begin{cases}
  \Tr[{A}\, P_{\phi_j^M}] \,\, & \text{if}  \,\, M=0 \,\, \text{and} \,\, j\leq 
d-1\\

0  \,\, & \text{if}  \,\, M=0 \,\, \text{and} \,\, j\geq d\\

\frac{d}{D}  \Tr[{A}\, P_{\phi_j^M}] \,\, & \text{if}  \,\, 1\leq M \leq (D+1) \, \\
& \text{and} \,\,0\leq j \leq D-1\, ,

 \end{cases}\nonumber
\end{equation}
in the same way for the operator $B$. Taking into 
account the fact that, by definition, ${\Tr\tilde{A} = \Tr A}$, ${\Tr\tilde{B} = \Tr B}$ and 
${\Tr[\tilde{A}\tilde{B}] = \Tr[AB]}$, we obtain:
\begin{equation}
 \sum_{\varphi \in X_d^D} c_{\varphi} \Tr[ A P_\varphi]\; \Tr[B P_\varphi] = 
\Tr A \,\Tr B+\Tr[AB] \, ,
\end{equation}
where $c_{\varphi} = 1$ if $\ket{\varphi}$ is one of the $d$ states with $M=0$, and 
$c_{\varphi} = d^2/D^2$ otherwise. Defining $Z \equiv \sum c_\varphi= d(d+1)$ and the probabilities 
$p_\varphi \equiv c_\varphi / Z$, we can rewrite the last equation as:
\begin{equation}
  \sum_{\varphi \in X_d} p_{\varphi} \bra{\varphi} A \ket{\varphi} \bra{\varphi} B \ket{\varphi} = 
\int_{\mathcal{H}_d} d\psi \bra{\psi} A \ket{\psi} \bra{\psi} B \ket{\psi} \, ,
\label{nonuniform2design}
\end{equation}
which shows that the set $X_d^D$ is indeed a proper $2$-design in 
$\mathcal{H}_d$. As we said before, unlike a complete set of mutually unbiased bases in prime power 
dimension, this set is not a uniform $2$-design. The set $X_d^D$ splits in two subsets whose weights are 
different when it comes to estimate Haar integrals. For the $d$ states with $M=0$ we have 
$p_\varphi = \frac{1}{Z}$, and for the remaining $D^2$ states $p_\varphi = \frac{d^2}{ZD^2}$. 

Finally, the application for SEQPT is direct: the protocol is exactly the 
same as the one  in Ref.~\cite{paper1conFer}, but the sampling of the states 
from the $2$-design $X_d^D$ is done according to the probability distribution $p_\varphi$. 
The efficiency of the protocol is also the same, and can also be obtained as the efficiency of the method in dimension $D$, as shown in \cite{paper1conFer}. That is, to obtain an error lower than $\epsilon$ with a probability of failure lower than $p$ one has to perform $M$ experiments, where $M\geq \ln \left(2/p\right)/2\epsilon^2$. The efficiency of the method comes from the fact that $M$ does not depend on the dimension $d$ (nor it depends on $D$).\newline

%xxxxxxxxxxxxxxxxxxxxxxxxxxxxxxxxxxxxxxxxxxxxxxxxxxxxxxxxxxxxxxxxxxxxxxxxxxxxxxxxxxxxxxxxxxxxxxxxxxxxxxxxxxxxxxxxxxxxxxxxxxxxxxxxxxxxxxxxxxxxxxxxxxxxxxxxxx
%xxxxxxxxxxxxxxxxxxxxxxxxxxxxxxxxxxxxxxxxxxxxxxxxxxxxxxxxxxxxxxxxxxxxxxxxxxxxxxxxxxxxxxxxxxxxxxxxxxxxxxxxxxxxxxxxxxxxxxxxxxxxxxxxxxxxxxxxxxxxxxxxxxxxxxxxxx

\section{Summary}
\label{Sec:disc}

Quantum process tomography is a key tool for quantum information processing. In general, full 
quantum tomography is an inefficient task and protocols for
 partial tomography provide essential information about a given process. In this article we 
showed two schemes that allow to perform efficient selective process tomography in arbitrary 
dimensions.
These protocols are based on the idea presented in Ref.~\cite{paper1conFer}, and are extended to 
arbitrary finite dimensions. Our approach requires the estimation of Haar integrals of quadratic 
forms
in dimensions where the construction of uniform 2-designs is not known. Thus, we showed how one can 
achieve this task using two methods: on one hand, an approximate 2-design can be constructed using 
tensor product of 2-designs and, on the other, a non-uniform 2-design can be obtained 
by projecting 2-designs over smaller dimensions.

\acknowledgments
We acknowledge financial support from ANPCyT (PICT 2014-3711 and PICT 2015-2293), CONICET and 
UBACyT.

\input{refs.tex}\begin{appendix}

\section{Estimating the elements of the $\chi$-matrix with tensor product of 2-designs}\label{app:tensor}

Here we sketch a derivation of Eq.~\eqref{eq:chifid}. In this case, the action of the channel $\E_{j_1j_2}^{i_1i_2}$ over
states of the 2-designs $X_1$ and $X_2$ is given by:
\bea
\E_{j_1j_2}^{i_1i_2}(P_{\psi_1}\otimes P_{\psi_2})&=& \sum_{\substack{\mu_1\nu_1 \\ \mu_2\nu_2}} \chi_{\nu_1\nu_2}^{\mu_1\mu_2} \left(E_{\mu_1}E^{i_1}P_{\psi_1}E_{j_1}E^{\nu_1}\right) \nonumber \\
&& \times \left(E_{\mu_2}E^{i_2}P_{\psi_2}E_{j_2}E^{\nu_2}\right).
\eea
In order to compute the average in this expression, 
we will consider that the initial states $P_{\psi_1}\otimes P_{\psi_2}$ are sampled from the set of states $X_\otimes$, i.e. the tensor product of 2-designs.

First, let us consider the average of the survival probability over the set $X_\otimes$. Using Eq.~\eqref{eq:meanfid} this average is given by:
\bea
\fid_\otimes(\E_{j_1j_2}^{i_1i_2}) &=& \sum_{\substack{\mu_1\nu_1 \\ \mu_2\nu_2}} \chi_{\nu_1\nu_2}^{\mu_1\mu_2}
\left(\frac{D_1^2\,\delta_{\mu_1}^{i_1}\delta_{j_1}^{\nu_1} +\Tr[E_{\mu_1}E^{i_1}E_{j_1}E^{\nu_1}]}{D_1(D_1+1)} \right)  \nonumber \\
&& \times \left(\frac{D_2^2\,\delta_{\mu_2}^{i_2}\delta_{j_2}^{\nu_2} +\Tr[E_{\mu_2}E^{i_2}E_{j_2}E^{\nu_2}]}{D_2(D_2+1)} \right). 
\label{eq:survp}
\eea

Now, let us consider the survival probability of system 1 independent of the result of the measurement over system 2. 
This mean value can be thought of as a reduced mean survival probability over system 1, and it can 
be evaluated just by tracing out system 2 in the circuit described in 
Sec.~\ref{Sec:seqptbip}.
Thus, using this idea and Eq.~\eqref{eq:meanfid} over system 1:
\bea
 \fid_1(\E_{j_1j_2}^{i_1i_2})&=& \sum_{\substack{\mu_1\nu_1 \\ \mu_2\nu_2}} \chi_{\nu_1\nu_2}^{\mu_1\mu_2}
\left(\frac{D_1^2\,\delta_{\mu_1}^{i_1}\delta_{j_1}^{\nu_1} +\Tr[E_{\mu_1}E^{i_1}E_{j_1}E^{\nu_1}]}{D_1(D_1+1)} \right)  \nonumber \\
&& \times \left(\sum_{\psi_2\in X_2}\frac{\Tr[E_{\mu_2}E^{i_2}P_{\psi_2}E_{j_2}E^{\nu_2}]}{D_2(D_2+1)} \right). \nonumber
\eea
Then, using that $\sum_{\psi_2\in X_2} P_{\psi_2}=(1+D_2) \Id_2$ and the trace preserving condition 
$\sum_{\substack{\mu_1\nu_1 \\ \mu_2\nu_2}} \chi_{\nu_1\nu_2}^{\mu_1\mu_2}E^{\nu_1\nu_2}E^{\mu_1\mu_2}=\Id$ we arrive at:
\bea
\fid_1(\E_{j_1j_2}^{i_1i_2})&=&\sum_{\mu_2\nu_2} \chi_{j_1\nu_2}^{i_1\mu_2}\, \Tr[E_{\mu_2}E^{i_2}E_{j_2}E^{\nu_2}] \frac{D_1}{D_2(D_1+1)} +\nonumber \\
&& + \frac{\delta_{j_1}^{i_1}\delta_{j_2}^{i_2}}{D_1+1}.
\eea
Similarly, we can obtain an analogous expression if we consider the reduced mean survival probability over system 2.
Finally, we can relate the mean fidelities by:
\beq
\fid_\otimes(\E_{j_1j_2}^{i_1i_2}) =  \frac{d \; \chi_{j_1j_2}^{i_1i_2}  - \delta_{j_1}^{i_1}\delta_{j_2}^{i_2}}{(D_1+1)(D_2+1)}+ \frac{\fid_1(\E_{j_1j_2}^{i_1i_2})}{D_2+1}+
\frac{\fid_2(\E_{j_1j_2}^{i_1i_2})}{D_1+1}
\eeq
We can also see explicitly the correction to the average over the Haar  given by the mean fidelity:
\bea
 \fid(\E_{j_1j_2}^{i_1i_2}) &=& \fid_\otimes(\E_{j_1j_2}^{i_1i_2}) \frac{(D_1+1)(D_2+1)}{d+1}  + \delta_{j_1}^{i_1}\delta_{j_2}^{i_2}\frac{2}{d+1}\nonumber \\
&& -\fid_1(\E_{j_1j_2}^{i_1i_2})\frac{(D_1+1)}{d+1}
-\fid_2(\E_{j_1j_2}^{i_1i_2})\frac{(D_2+1)}{d+1}.\nonumber
\eea

\section{Efficiency}\label{app:efficiency}

In the bipartite case, the circuit of Fig.~\ref{seqpt2primos} can give eight different results: 
they are of 
the form ${\pm\vec{r}}$ where the first symbol ($\pm$) refers to the polarization obtained for the 
auxiliary qubit, and $\vec{r}$ is a binary string of length two such that $r_i=1$ if the projective 
measurement on subsystem $\mathcal{H}_i$ gives survival of the initial state and $r_i=0$ if not. If 
we perform $M$ runs of the experiment (where each run corresponds to choosing the initial states 
randomly from the $2$-designs $X_1$ and $X_2$), we define $M_{\pm\vec{r}}$ as the number of times 
we obtain result ${\pm\vec{r}}$. So, for example, $M_{-01}$ is the number of times we measure $-1$ 
for the polarization of the auxiliary qubit, non-survival of the initial state on $\mathcal{H}_1$ 
and survival of the initial state on $\mathcal{H}_2$. Having said that, if we perform $M$ runs of 
the experiment, estimations of $\bar F_\otimes(\mathcal E_{j_1j_2}^{i_1i_2})$, $\bar 
F_1(\mathcal E_{j_1j_2}^{i_1i_2})$ and $\bar F_2(\mathcal E_{j_1j_2}^{i_1i_2})$ will be given by:
\bea
&&\bar F_\otimes(\mathcal 
E_{j_1j_2}^{i_1i_2}) \simeq \frac{1}{M} \left( M_{+11} - M_{-11} \right)    \nonumber \\
&&\bar F_1(\mathcal 
E_{j_1j_2}^{i_1i_2}) \simeq \frac{1}{M} \left( M_{+11} + M_{+10} - M_{-11} - M_{-10} \right)    
 \\
&&\bar F_2(\mathcal 
E_{j_1j_2}^{i_1i_2}) \simeq \frac{1}{M} \left( M_{+11} + M_{+01} - M_{-11} - M_{-01} \right) \, .
\nonumber
\eea
Using this in Eq.~(\ref{eq:chifid}) we obtain an expression for the estimation of 
$\chi_{j_1j_2}^{i_1i_2}$:
\bea
\chi_{j_1j_2}^{i_1i_2} \simeq  \frac{1}{M} && \left\{  \left( 1- \frac{1}{d} \right) \left[ M_{+11} 
- M_{-11} \right] \right. \nonumber \\
&& - \left( \frac{D_1+1}{d} \right) \left[ M_{+10} - M_{-10} \right] \\
&& \left. - \left( \frac{D_2+1}{d} \right) \left[ M_{+01} - M_{-01} \right] \right\} \, ,\nonumber
\eea
so we can think of $\chi_{j_1j_2}^{i_1i_2}$ as the average of a random variable which can 
take values $\pm   \left( 1- \frac{1}{d} \right)$, $\pm \left( \frac{D_1+1}{d} \right)$, 
$\pm \left( \frac{D_2+1}{d} \right)$  and $0$ (corresponding to the experiments where none of the 
initial states survive). Hoeffding's inequality \cite{hoeffding} tells us that if $x$ is a random 
variable that takes values in the interval $[a,b]$ and $\overline x = \frac{1}{n} \sum_i^n x_i$ is 
the average of $n$ realizations of this variable, then:
\begin{equation}
\textrm{Prob} \left(\left |\overline x - \mathrm{E} [ x ] \right | \geq \epsilon \,  \right) \leq 2\exp \left(\frac{-2n\epsilon^2}{(b - a)^2} \right) \, ,
\end{equation}
where $\mathrm{E}[x]$ is the mean value of 
$x$. In our case, $\chi_{j_1j_2}^{i_1i_2}$ is estimated as the mean value of $M$ realizations of a 
random variable which takes values in $\left[ -  \left( 1- \frac{1}{d} \right) ,  \left( 1- 
\frac{1}{d} \right) \right]$, so the probability $p$ of failure when estimating this coefficient 
with error $\varepsilon$ is bounded as:
\begin{equation}
 p \leq 2 \exp \left( \frac{-M\varepsilon ^2}{2(1-\frac{1}{d})^2} \right) \, ,
\end{equation}
which is the same as in Eq.~(\ref{efficiency2}).

\begin{widetext}
In the general case of $N$ subsystems, a similar reasoning gives 
\begin{equation}
 \chi_{j}^{i} \simeq \frac{1}{M} \sum_{\vec{r}} C_{\vec{r}} \left( M_{+\vec{r}} - 
M_{-\vec{r}} \right) \, ,
\end{equation}
where now $\vec{r}$ is a binary string of length $N$ indicating survival (or not) in each 
subsystem. The coefficients $C_{\vec{r}}$ are:
\begin{equation}
  C_{\vec{r}} = \begin{cases}
                           0 \,\, & \text{if}  \,\, \vec{r}=\vec{0}\\
\frac{1}{d} \left[ \prod_{a=1}^N (D_a+1) - \sum_{\vec{m} \notin  \{\vec{0},\vec{1}\} }  
\prod_{a/\vec{m}_a =1} (D_a+1) \right] \,\, & 
\text{if}  \,\, \vec{r} = \vec{1} \,  \\
-\frac{1}{d} \sum_{\vec{m} \notin \{\vec{0},\vec{1}\}  / \vec{m} \cdot \vec{r} = \left \| \vec{m} 
\right \|_1 } \prod_{a/\vec{m}_a =1} (D_a+1) \,\, & 
\text{otherwise} \, ,
\end{cases}
\end{equation}
\end{widetext}
where $\left \| \vec{w}  \right \|_1$ is the sum of the elements of $\vec{w}$. From the last 
equation, it is easy to get a useful bound for those coefficients: $|C_{\vec{r}}| < 
4^N$. Hence, in general, we can think of $\chi_{j}^{i}$ as the mean value of a 
random variable in $[-4^N,4^N]$ and Hoeffding's inequality gives:
\begin{equation}
 p \leq 2 \exp \left( \frac{-M\varepsilon ^2}{2\times4^N} \right) \, ,
\end{equation}
where $p$ is the failure probability for the estimation of $\chi_{j}^{i}$ with 
error $\varepsilon$ using the results of $M$ experiments. From there, it is easy to show that
\begin{equation}
 M \geq \frac{2 \log(2/p) 4^N }{\varepsilon ^2}
 \label{experimentos}
\end{equation}
experiments are enough to achieve the desired precision. By definition, $N=\omega(d)$ 
(the number of distinct prime divisors of $d$), and the Hardy–Ramanujan theorem \cite{ramanujan} 
states that:
\begin{equation}
 |\omega(d)-\log(\log(d))|<{(\log(\log(d)))}^{\frac12 +\delta}
 \label{HRtheorem}
\end{equation}
for almost all $d\in \mathbb{N}$ and for all $\delta > 0$. In particular, 
$\omega(d) < \log(\log(d)) + (\log(\log(d)))^{\frac{1}{2} +\delta}$. Picking $\delta = 1/2$, we get:
\begin{equation}
 \omega ( d ) < 2 \log (\log(d))
\end{equation}
for almost all $d\in \mathbb{N}$. Using this in Eq.~(\ref{experimentos}), and recalling that $N= 
\omega(d)$, we obtain that
\begin{equation}
 M \geq \frac{2 \log (2/p) \log(d) ^4 }{\varepsilon ^ 2}
\end{equation}
experiments are enough for almost all dimensions $d$. Thus, the efficiency of the protocol follows from 
the fact that the number of experiments grows polynomially with the logarithm of the dimension.

\end{appendix}

\end{document}

%% file: macros.tex
\newcommand{\braket}[2]{\left\langle #1 \vert #2\right\rangle}

\usepackage{amsmath}
\usepackage{bbm,amsfonts}

\def\Id{{\mathbbm 1}}

\def\E{{\mathcal E}}

\usepackage{amsthm}
\theoremstyle{definition}

%% file: TP2D.bbl
\begin{thebibliography}{10}
\expandafter\ifx\csname url\endcsname\relax
  \def\url#1{\texttt{#1}}\fi
\expandafter\ifx\csname urlprefix\endcsname\relax\def\urlprefix{URL }\fi
\providecommand{\bibinfo}[2]{#2}
\providecommand{\eprint}[2][]{\url{#2}}

\bibitem{paper1conFer}
\bibinfo{author}{Bendersky, A.}, \bibinfo{author}{Pastawski, F.} \&
  \bibinfo{author}{Paz, J.~P.}
\newblock \bibinfo{title}{Selective and efficient estimation of parameters for
  quantum process tomography}.
\newblock \emph{\bibinfo{journal}{Phys. Rev. Lett.}}
  \textbf{\bibinfo{volume}{100}}, \bibinfo{pages}{190403}
  (\bibinfo{year}{2008}).

\bibitem{paper2conFer}
\bibinfo{author}{Bendersky, A.}, \bibinfo{author}{Pastawski, F.} \&
  \bibinfo{author}{Paz, J.~P.}
\newblock \bibinfo{title}{Selective and efficient quantum process tomography}.
\newblock \emph{\bibinfo{journal}{Phys. Rev. A}} \textbf{\bibinfo{volume}{80}},
  \bibinfo{pages}{032116} (\bibinfo{year}{2009}).

\bibitem{NielsenChuang}
\bibinfo{author}{Nielsen, M.~A.} \& \bibinfo{author}{Chuang, I.~L.}
\newblock \emph{\bibinfo{title}{{Quantum Computation and Quantum Information
  (Cambridge Series on Information and the Natural Sciences)}}}
  (\bibinfo{publisher}{Cambridge University Press}, \bibinfo{year}{2004}),
  \bibinfo{edition}{1} edn.
\newblock \urlprefix\url{http://www.worldcat.org/isbn/521635039}.

\bibitem{DCQD}
\bibinfo{author}{Mohseni, M.} \& \bibinfo{author}{Lidar, D.~A.}
\newblock \bibinfo{title}{Direct characterization of quantum dynamics}.
\newblock \emph{\bibinfo{journal}{Phys. Rev. Lett.}}
  \textbf{\bibinfo{volume}{97}}, \bibinfo{pages}{170501}
  (\bibinfo{year}{2006}).

\bibitem{MohsRezLid}
\bibinfo{author}{Mohseni, M.}, \bibinfo{author}{Rezakhani, A.~T.} \&
  \bibinfo{author}{Lidar, D.~A.}
\newblock \bibinfo{title}{Quantum-process tomography: Resource analysis of
  different strategies}.
\newblock \emph{\bibinfo{journal}{Phys. Rev. A}} \textbf{\bibinfo{volume}{77}},
  \bibinfo{pages}{032322} (\bibinfo{year}{2008}).

\bibitem{paperConCeciLopez}
\bibinfo{author}{L\'opez, C.~C.}, \bibinfo{author}{Bendersky, A.},
  \bibinfo{author}{Paz, J.~P.} \& \bibinfo{author}{Cory, D.~G.}
\newblock \bibinfo{title}{Progress toward scalable tomography of quantum maps
  using twirling-based methods and information hierarchies}.
\newblock \emph{\bibinfo{journal}{Phys. Rev. A}} \textbf{\bibinfo{volume}{81}},
  \bibinfo{pages}{062113} (\bibinfo{year}{2010}).

\bibitem{prlChris}
\bibinfo{author}{Schmiegelow, C.~T.}, \bibinfo{author}{Bendersky, A.},
  \bibinfo{author}{Larotonda, M.~A.} \& \bibinfo{author}{Paz, J.~P.}
\newblock \bibinfo{title}{Selective and efficient quantum process tomography
  without ancilla}.
\newblock \emph{\bibinfo{journal}{Phys. Rev. Lett.}}
  \textbf{\bibinfo{volume}{107}}, \bibinfo{pages}{100502}
  (\bibinfo{year}{2011}).
\newblock
  \urlprefix\url{http://link.aps.org/doi/10.1103/PhysRevLett.107.100502}.

\bibitem{SCNQP}
\bibinfo{author}{Emerson, J.} \emph{et~al.}
\newblock \bibinfo{title}{{Symmetrized Characterization of Noisy Quantum
  Processes}}.
\newblock \emph{\bibinfo{journal}{Science}} \textbf{\bibinfo{volume}{317}},
  \bibinfo{pages}{1893--1896} (\bibinfo{year}{2007}).
\newblock \urlprefix\url{http://dx.doi.org/10.1126/science.1145699}.

\bibitem{schwinger1960unitary}
\bibinfo{author}{Schwinger, J.}
\newblock \bibinfo{title}{Unitary operator bases}.
\newblock \emph{\bibinfo{journal}{Proceedings of the National Academy of
  Sciences}} \textbf{\bibinfo{volume}{46}}, \bibinfo{pages}{570--579}
  (\bibinfo{year}{1960}).

\bibitem{ivanovic1981geometrical}
\bibinfo{author}{Ivanovic, I.}
\newblock \bibinfo{title}{Geometrical description of quantal state
  determination}.
\newblock \emph{\bibinfo{journal}{Journal of Physics A: Mathematical and
  General}} \textbf{\bibinfo{volume}{14}}, \bibinfo{pages}{3241}
  (\bibinfo{year}{1981}).

\bibitem{wootters1989optimal}
\bibinfo{author}{Wootters, W.~K.} \& \bibinfo{author}{Fields, B.~D.}
\newblock \bibinfo{title}{Optimal state-determination by mutually unbiased
  measurements}.
\newblock \emph{\bibinfo{journal}{Annals of Physics}}
  \textbf{\bibinfo{volume}{191}}, \bibinfo{pages}{363--381}
  (\bibinfo{year}{1989}).

\bibitem{durt2010mutually}
\bibinfo{author}{Durt, T.}, \bibinfo{author}{Englert, B.-G.},
  \bibinfo{author}{Bengtsson, I.} \& \bibinfo{author}{{\.Z}yczkowski, K.}
\newblock \bibinfo{title}{On mutually unbiased bases}.
\newblock \emph{\bibinfo{journal}{International journal of quantum
  information}} \textbf{\bibinfo{volume}{8}}, \bibinfo{pages}{535--640}
  (\bibinfo{year}{2010}).

\bibitem{Ambainis07}
\bibinfo{author}{Ambainis, A.} \& \bibinfo{author}{Emerson, J.}
\newblock \bibinfo{title}{Quantum t-designs: t-wise independence in the quantum
  world.}
\newblock In \emph{\bibinfo{booktitle}{IEEE Conference on Computational
  Complexity'07}}, \bibinfo{pages}{129--140} (\bibinfo{year}{2007}).

\bibitem{klappenecker2005mutually}
\bibinfo{author}{Klappenecker, A.} \& \bibinfo{author}{Rotteler, M.}
\newblock \bibinfo{title}{Mutually unbiased bases are complex projective
  2-designs}.
\newblock In \emph{\bibinfo{booktitle}{Information Theory, 2005. ISIT 2005.
  Proceedings. International Symposium on}}, \bibinfo{pages}{1740--1744}
  (\bibinfo{organization}{IEEE}, \bibinfo{year}{2005}).

\bibitem{dankert-2005}
\bibinfo{author}{Dankert, C.}
\newblock \emph{\bibinfo{title}{{Efficient Simulation of Random Quantum States
  and Operators}}}.
\newblock Master's thesis, \bibinfo{school}{University of Waterloo},
  \bibinfo{address}{Canada} (\bibinfo{year}{2005}).

\bibitem{hoeffding}
\bibinfo{author}{Hoeffding, W.}
\newblock \bibinfo{title}{Probability inequalities for sums of bounded random
  variables}.
\newblock \emph{\bibinfo{journal}{Journal of the American Statistical
  Association}} \textbf{\bibinfo{volume}{58}}, \bibinfo{pages}{13--30}
  (\bibinfo{year}{1963}).

\bibitem{ramanujan}
\bibinfo{author}{{Hardy}, G.~H.} \& \bibinfo{author}{{Ramanujan}, S.}
\newblock \bibinfo{title}{{The normal number of prime factors of a number
  $n$.}}
\newblock \emph{\bibinfo{journal}{{Quart. J.}}} \textbf{\bibinfo{volume}{48}},
  \bibinfo{pages}{76--92} (\bibinfo{year}{1917}).

\end{thebibliography}
